\documentclass[12pt]{article}
\usepackage{graphicx}
\usepackage[cp1251]{inputenc}
\usepackage{rotating}

\begin{document}
 \noindent {\footnotesize\it
   Astronomy Letters, 2022, Vol. 48, No. 10, pp. 568--577}
 \newcommand{\dif}{\textrm{d}}

 \noindent
 \begin{tabular}{llllllllllllllllllllllllllllllllllllllllllllll}
 & & & & & & & & & & & & & & & & & & & & & & & & & & & & & & & & & & & & & &\\\hline\hline
 \end{tabular}

  \vskip 0.5cm
  \bigskip
 \bigskip
 \centerline{\bf Redetermination of the Parameters of the Galactic Spiral Pattern}
 \centerline{\bf from Classical Cepheids}

 \bigskip
 \bigskip
  \centerline { %%$\copyright$ 2022 г. \qquad
   V. V. Bobylev\footnote [1]{vbobylev@gaoran.ru},  A. T. Bajkova}
 \bigskip
 \centerline{\small\it Pulkovo Astronomical Observatory, Russian Academy of Sciences, St. Petersburg, 196140 Russia}
 \bigskip
 \bigskip
 {
A sample of classical Cepheids of the Galaxy with estimates of their distances taken from the work of Skowron et al., where they were determined on the basis of the period-luminosity relation has been studied. In the present work, the distances of Skowron et al. were increased by 10\% according to the results of our previous kinematic analysis of these Cepheids. Geometric characteristics of two spiral arms, namely, the Carina-Sagittarius and the Outer arms have been specified. The distance from the Sun to the galactic center was assumed to be $8.1\pm0.1$~kpc.
Based on 257 Cepheids belonging to a segment of the Carina-Sagittarius arm, with ages in the range of 80--120 Myr, the value of the pitch angle of the spiral pattern $i=-12.7\pm0.4^\circ$ and the position of this arm $a_0=7.28\pm0.05$~kpc are found. Based on 352 Cepheids from the Outer arm with ages in the range of 120--300 Myr there were found the estimates: $i=-12.0\pm0.5^\circ$ and $a_0=13.03\pm0.06$~kpc. Based on a sample of 1618 Cepheids with ages in the range of 80–300 Myr, a wavelet map was constructed in the ``position angle--logarithm of distance'' plane.
From the analysis of this map, the following estimates were obtained for the Carina-Sagittarius arm: $i=-12.9\pm0.4^\circ$ and $a_0=7.43\pm0.05$~kpc, and for the Outer arm $i=-12.5\pm0 .5^\circ$ and $a_0=13.33\pm0.06$~kpc.
 }

\bigskip
%\noindent
%{\it Key words:} classical Cepheids, spiral pattern, structure of the Galaxy.
%\newpage

\section*{INTRODUCTION}
A huge number of publications have been devoted to the study of the spiral structure of the Galaxy using various objects and methods (for example, Lin, and Shu, 1964; Lin et al., 1969; Cr\'ez\'e, and Mennessier, 1973; Mishurov et al. 1979; Mishurov, and Zenina, 1999; L\'epine et al., 2001; Moitinho et al., 2006; V\'azquez et al., 2008; Gerhard, 2011; Efremov, 2011; Hou, and Han, 2014; 2015; Poggio et al., 2021). Various stars are used as indicators of the spiral structure, for example, Cepheids (Mishurov et al., 1979; Mishurov, and Zenina, 1999; Dambis et al., 2015) or OB stars (Xu et al.,,2018; 2021), open star clusters (Popova, and Loktin, 2005; Dias, and L\'epine, 2005; Hao et al., 2021) and OB associations (Mel'nik et al., 2001), clouds of neutral (Levine et al., 2006) and ionized hydrogen (Paladini et al., 2004), clouds of interstellar dust (Taylor, and Cordes, 1993), star-forming regions (Georgelin, and Georgelin, 1976; Russeil, 2003), sources of maser emission (Bobylev, and Baikova, 2014; Reid et al., 2019), etc. Methods are used based both on the analysis of the spatial distribution of stars using the measured distances to them, and on the study of clusters of objects distributed along the galactic equator.

Until now, there is no generally accepted model of the global (grand design) spiral structure of the Galaxy. Models with a different number of spiral arms, with a constant or variable pitch angle, symmetric or asymmetric spirals are discussed. Modern data on the distribution of clouds of neutral and ionized hydrogen, as well as maser sources with trigonometric parallaxes measured by the VLBI method, rather speak of a four-arm model with a constant value of the pitch angle in the interval of 10--14$^\circ$. A large evidence base in favor of the four-armed global pattern is collected in reviews by Vall\'ee (1995; 2002; 2008; 2017).

Some authors prefer to analyze not the global structure, but individual segments of spiral arms, close to the Sun, with individual inclination angles (Nikiforov, Veselova,
2018; Veselova, Nikiforov, 2020). Estimates of pitch angles found from individual segments of spiral arms are in the range 9--18$^\circ$ (Griv et al., 2017; Reid et al., 2019; Hao et al., 2021). A discussion of the modern observed picture associated with the spiral structure of the Galaxy can be found, for example, in the reviews by Hou, and Han~(2014), Vall\'ee~(2017), Xu et al. (2018) or Hou~(2021).

It is well known that classical Cepheids trace a spiral pattern. Although such stars are not representatives of the youngest population, nevertheless, they are of great value for studying the spiral structure of the Galaxy. This is possible due to the high accuracy of estimating the distance to them using the period-luminosity relation (with errors of 5--10\%) and their distribution over a very wide region of the Galaxy.

Bobylev (2022) found the parameters of the galactic spiral pattern from a large sample of classical Cepheids. The distances to these Cepheids were calculated from the period-luminosity relationship by Skowron et al.~(2019) using mid-infrared photometry. However, Bobylev, and Bajkova (2022), based on a kinematic analysis of Cepheids, as well as a direct comparison of various distance scales, showed the need to extend the distance scale of Skowron et al.~(2019) by about 10\%.

In this paper, we actually repeat the analysis of Bobylev (2022), taking into account the fact that we use the distances to Cepheids from Skowron et al.~(2019) increased by 10\%. Thus, the goal of this work is to redefine the parameters of the spiral pattern of Galaxies from a sample of classical Cepheids with distances extended by 10\%.

 %\newpage
 \section*{METHODS}\label{method}
The position of a star in a logarithmic spiral wave can be described by the following equation:
 \begin{equation}
 R=R_0 e^{(\theta-\theta_0)\tan i},
 \label{spiral-1}
 \end{equation}
where $R$ is a distance from the center of the Galaxy to the star; $R_0$ is a distance from the center of the Galaxy to the Sun; $\theta$ is a position angle of the star: $\tan\theta=y/(R_0-x)$, where $x,y$ are heliocentric galactic rectangular coordinates of the star, with the $x$ axis directed from of the Sun to the galactic center, and the direction of the $y$ axis coincides with the direction of the galactic rotation; $\theta_0$ is some arbitrarily chosen initial angle; $i$ is a pitch angle of the spiral pattern ($i<0$ for a twisting spiral).
In this paper, the value of $R_0$ is taken equal to $8.1\pm0.1$~kpc according to survey of Bobylev, and Bajkova (2021), where it was derived as a weighted average from a large number of modern individual estimates.

Since $\theta_0$ is a constant, and the approximate value of $\tan i$ is known to us from previous studies, we can take $\theta_0\tan i={\rm const}$ as a first approximation. Now the equation~(\ref{spiral-1}) can be rewritten as follows:
\begin{equation}
  \ln (R/R_0)=\theta\tan i+{\rm const},
 \label{spiral-02}
\end{equation}
or in a more convenient form
\begin{equation}
  \ln (R/R_0)=a\theta+b.
 \label{spiral-03}
\end{equation}
As you can see, the relation~(\ref{spiral-03}) is the equation of a straight line in the plane ``position angle~--logarithm of distance''. Solving the system of conditional equations separately for each segment of the spiral arm using the least squares method (LSM), we can find two quantities: $a$ and $b$. It is obvious that $a=\tan i$. Now suppose that $\theta=0,$ then we find the value $a_0=R_0e^b$~--- the place where the center of the considered spiral arm intersects the axis $X,$ directed from the center of the Galaxy and passing through the Sun. That is, the parameter $a_0$ specifies the radial position of the center of the spiral arm on the $X$ axis. Finally, note that in this method the estimate of the pitch angle $i$ does not depend on the number of spiral arms $m$.

In the first method, we exactly repeat the approach implemented in the work of Bobylev (2022), where it is shown that the geometric characteristics of two spiral arms, namely, the Carina-Sagittarius and the Outer arms are confidently determined from the Cepheids from Skowron et al. (2019). It is also shown that relatively young Cepheids with ages in the range of 80--120 Myr belong to the segment of the Carina-Sagittarius arm, while older Cepheids with ages in the range of 120--300 Myr belong to the segment of the Outer arm. With this approach, the selection of Cepheids necessary for analysis, belonging to segments of the Carina-Sagittarius and the Outer spiral arms, is carried out using pre-selection zones.

In the other method, we do not use predefined preselection zones. In order to extract statistically significant signals of the main inhomogeneities in the considered distribution ($\theta,\ln (R/R_0)$), we use the wavelet transform, which is known as a powerful tool for filtering spatially localized signals (Chui, 1997).

The wavelet transform of the two-dimensional distribution $f(X,Y)$ consists in its expansion in terms of analyzing wavelet functions $\psi(X/c,Y/c)$, where $c$ is a coefficient that allows one to select a function of a certain scale from the entire family of functions characterized by the same form $\psi$. The wavelet transform $\omega(\xi,\eta)$ is defined as a correlation function in such a way that at any given point $(\xi,\eta)$ on the $(X,Y)$~--- plane we have one real value of the following integral:
\begin{equation}
 \renewcommand{\arraystretch}{2.4}
   \displaystyle
 \omega(\xi,\eta)=\int_{-\infty}^{\infty}\int_{-\infty}^{\infty}f(X,Y)
 \psi\Biggl(\frac{(X-\xi)}{c},\frac{(Y-\eta)}{c}\Biggr)dX dY,
 \label{wav-01}
\end{equation}
which is called the wavelet coefficient at the point $(\xi,\eta)$. Obviously, in the case of finite discrete maps with which we operate, their number is finite and equal to the number of bins on the map.

We use a traditional wavelet function called the Mexican hat~--- Mexican HAT (MHAT). The two-dimensional MHAT function is described by the expression
\begin{equation}
 \renewcommand{\arraystretch}{2.0}
\psi(d/c)=\Biggl(2-\frac{d^2}{c^2}\Biggr)\exp^{-d^2/2c^2},
\label{wav-02}
\end{equation}
where $d^2=X^2+Y^2$.

The wavelet (\ref{wav-02}) is obtained as a result of a twofold differentiation of the Gaussian function. The main property of the $\psi$ function is that its integral over $X$ and $Y$ is equal to zero, which makes it possible to detect any irregularities in the distribution under study. If the analyzed distribution is uniform, then all wavelet transform coefficients will be equal to zero. Since the spiral density wave can be considered as a distribution of inhomogeneities in the plane of the Galaxy, the application of wavelet analysis to determine its characteristics seems justified and very interesting.

 \section*{DATA}\label{data}
The main source of data on classical Cepheids in this work is the catalog of Skowron et al. (2019). These Cepheids were observed within the fourth stage of the OGLE program (Optical Gravitational Lensing Experiment, Udalsky et al., 2015). This catalog contains estimates of the distance, age, period of pulsation, and photometric characteristics of Cepheids.

Heliocentric distances of 2214 Cepheids were calculated by Skowron et al. (2019) based on the period-luminosity relationship. Moreover, the specific ratio was taken by them from the work of Wang et al.~(2018), where it was refined from the light curves of Cepheids in the mid-infrared range. Skowron et al. (2019) estimated the age using the method of Anderson et al. (2016), taking into account the period of axial rotation of stars and the metallicity index.

To solve the problem posed in this paper, a sample of stars is required, the distances to which are determined using a single calibration from homogeneous data. Therefore, we do not add to the Cepheids from the list of Skowron et al. (2019) data on other known Cepheids, the distances to which are determined by other authors.

In this paper, we use the distances to Cepheids from Skowron et al.~(2019) increased by 10\%. The conclusion about the need for such an extension of the distance scale by Skowron et al. (2019) was made by Bobylev, and Bajkova (2022) during a kinematic analysis of a sample of 363 Cepheids younger than 120 Myr.

%%%%%%%%%%%%%%%%%%%%%%%%%%%%%%%%%%%%%%%%%%%%%%%%% FIG.1:
\begin{figure}[t]
{ \begin{center}
  \includegraphics[width=0.95\textwidth]{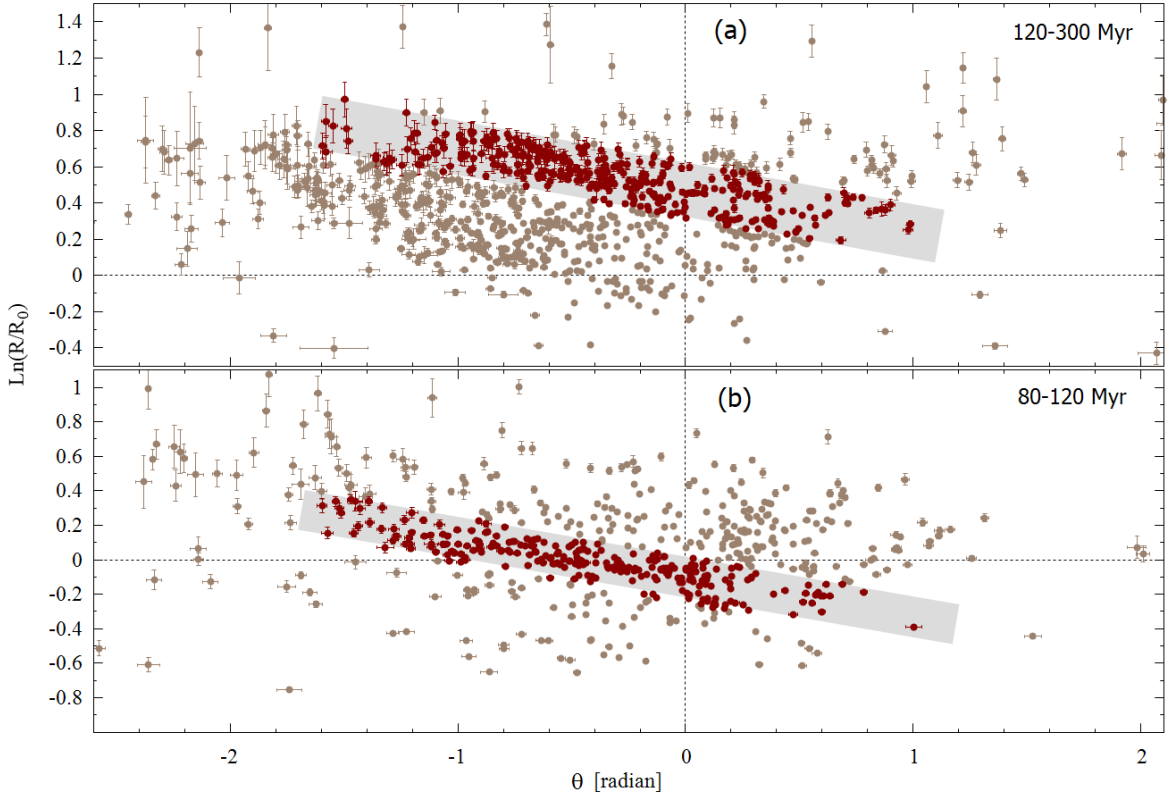}
  \caption{
``Position angle--logarithm of distance'' diagram for Cepheids with ages in the interval 120--300 Myr (a) and in the interval 80--120 Myr (b), the shading in each graph shows the region of the preliminary selection of stars.
}
 \label{f-lnR}
\end{center}}
\end{figure}
%%%%%%%%%%%%% Fig.1

 \section*{RESULTS AND DISCUSSION}
 \subsection*{Use of pre-screening zones}
Figure~\ref{f-lnR} shows the ``position angle--logarithm of distance'' diagrams for two samples of Cepheids. Boundaries are shown for the preliminary selection of Cepheids with ages in the range of 80--120 Myr, where the average age is about 100 Myr, and in the range of 120--300 Myr, with an average age of about 200 Myr.

The selection of Cepheids necessary for analysis is carried out using the same restrictions as in Bobylev (2022). So, for the stars in the segment of the Carina-Sagittarius arm, we used
the following limits: $-1.7<\theta<1.1$~rad, and $\ln(R/R_0)$ is bounded by two lines above and below $\theta\cdot\tan(-13^\circ)+0.02<\ ln(R/R_0)<\theta\cdot\tan(-13^\circ)-0.25$. Thus, here the total width of the sampling zone (from the inner to the outer edge of the arm along the $X$ axis) is 2.51 kpc.

To select Cepheids belonging to the segment of the Outer arm, the angle $\theta$ was taken from interval $-1.6<\theta<1.1$~rad., and the following restrictions for $\ln(R/R_0)$ are used:
$\theta\cdot\tan(-13^\circ)+0.6208<\ln(R/R_0)<\theta\cdot\tan(-13^\circ)+0.3120$. Here, the total width of the sampling zone is 2.93 kpc. The pitch angle value $i=-13^\circ$ was taken according to Bobylev, and Bajkova (2014), where it was estimated from a sample of masers with measured trigonometric parallaxes using the VLBI method.

From an analysis of the distribution of masers with trigonometric parallaxes measured by the VLBI method, Reid et al. (2019) obtained an estimate for the width parameter of the Carina-Sagittarius arm $1\sigma=0.27\pm0.05$~kpc, and the width of the Outer arm~--- $1\sigma=0.65\pm0.16$~kpc, where this parameter is calculated as the variance of the deviation from the dependence~(\ref{spiral-03}). Thus, according to Reid et al. (2019), the total width (from the inner to the outer edge of the arm along the $X$ axis, covering 98\% of the stars belonging to the arm) of the Carina-Sagittarius arm will be $6\sigma=1.62$~kpc, and of the Outer arm $6\sigma=3.90$~kpc. We can see that in our case, the selection zone width in the case of the Carina-Sagittarius arm is taken with a large margin, and in the case of the Outer arm, it is less than the estimate proposed by Reid et al.(2019). It is obvious that the value of $6\sigma$ should not exceed the wavelength of the spiral pattern $\lambda$ (to avoid overlapping of the arms). For example, in the region of the Sun $\lambda\sim3$~kpc, in a logarithmic wave this value increases with increasing galactocentric distance, therefore, in the region of the Outer arm $\lambda\sim4$~kpc.

Table~\ref{t1} gives the parameters $a$ and $b$ found as a result of the least squares solution of the system of conditional equations~(\ref{spiral-03}). The search for a solution was carried out both with unit weights and with weight coefficients $w_i=1/\sigma^2_{\ln(R/R_0)}$, where $i=1,...,n_\star$, and $ n_\star$~--- the number of stars used in the solution.

%%%%%%%%%%%%%%%%%%%%%%%%%%%%%%%%%%%%%
  \begin{table*}
   \begin{center}
    \caption{
The parameters of the spiral, found when searching for the least squares solution of a system of conditional equations~(\ref{spiral-03}) with unit weights, are given in the upper part of the table, and in the lower part with weights of the form $w_i=1/\sigma^ 2_{\ln(R/R_0)}$, $n_\star$~--- number of used Cepheids
    }
   \label{t1}
   {\small
   \begin{tabular}{|l|c|c|c|r|r|}      \hline
 Arm          & $n_\star$ & $a$ & $b$ & $i,$ deg. & $a_0$, kpc\\\hline
 II (Carina-Sagittarius)&257&$-0.222\pm0.008$& $-0.105\pm0.005$& $-12.51\pm0.45$&$ 7.29\pm0.05$\\
 IV  (Outer)     &352&$-0.218\pm0.008$& $+0.474\pm0.005$& $-12.31\pm0.48$&$13.01\pm0.06$\\
 \hline
 II (Carina-Sagittarius)&257&$-0.225\pm0.008$& $-0.106\pm0.005$& $-12.67\pm0.44$&$ 7.28\pm0.05$\\
 IV  (Outer)     &352&$-0.212\pm0.008$& $+0.483\pm0.006$& $-11.96\pm0.46$&$13.03\pm0.06$\\
 \hline
      \end{tabular}}
     \end{center}
   \end{table*}
%%%%%%%%%%%%%%%%%%%%%%%%%%%%%%%%%%%%%

For comparison, Bobylev (2022) obtained the following estimates for 269 Cepheids belonging to the segment of the Carina-Sagittarius arm using unit weights: $i=-12.0\pm0.5^\circ$ and $a_0=7.29\pm0. 05$~kpc. Similarly, for 343 Cepheids from the Outer arm: $i=-11.7\pm0.5^\circ$ and $a_0=12.81\pm0.06$~kpc. Calculations performed with weights of the form $w_i=1/\sigma^2_{\ln(R/R_0)},$ made it possible to find for the Carina-Sagittarius arm: $i=-11.9\pm0.2^\circ$ and $a_0 =7.32\pm0.05$~kpc, and for the Outer arm $i=-11.5\pm0.5^\circ$ and $a_0=12.89\pm0.06$~kpc. We can see that the lengthening of the Cepheid distance scale by 10\% changed the values ??of the required parameters $i$ and $a_0$ approximately by $1\sigma$.

As in work of Reid et al. (2019), we calculated the variance from the deviation from the dependence~(\ref{spiral-03}). As a result, $1\sigma=0.58$~kpc was found for the Carina-Sagittarius arm, and $1\sigma=0.72$~kpc for the Outer arm. For the Outer arm, our estimate differs little from that of Reid et al. (2019). But for the Carina-Sagittarius arm, our estimate is approximately twice that of Reid et al. (2019). This difference is most likely due to the fact that the Carina-Sagittarius arm in our case is traced by Cepheids with ages in the range of 80--120 Myr, which have managed to significantly move away from their birthplace. This leads to stronger scattering from the center of the spiral arm compared to the very young masers used by Reid et al. (2019).

%%%%%%%%%%%%%%%%%%%%%%%%%%%%%%%%%%%%%%%%%%%%%%%%% FIG.2:
\begin{figure}[t]
{ \begin{center}
  \includegraphics[width=0.98\textwidth]{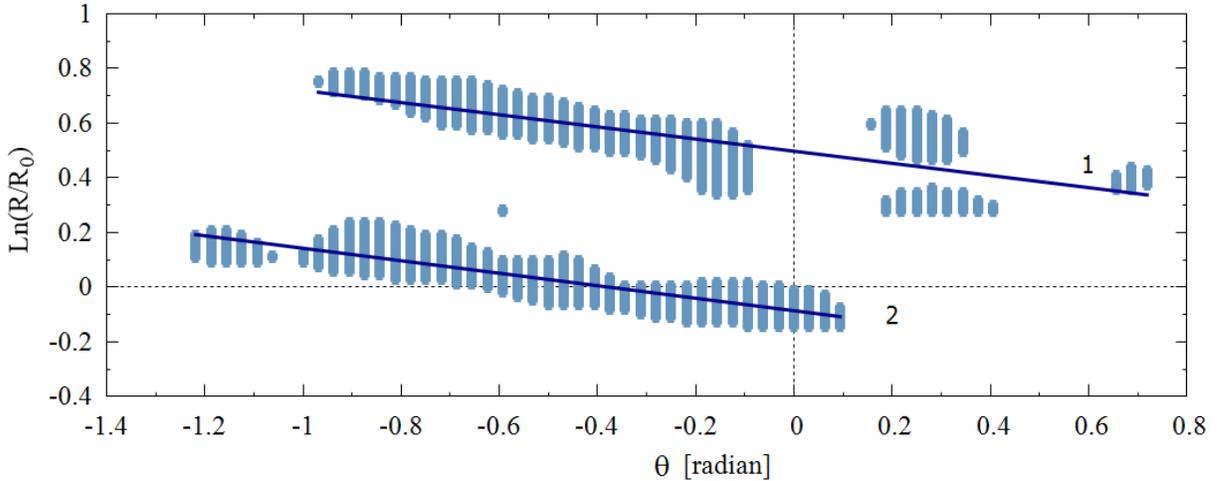}
  \caption{
Wavelet map in the diagram ``position angle--logarithm of distance'', constructed from a combined sample of Cepheids with ages in the range of 80--300 Ma, line 1 is drawn along the Outer arm, and line 2 along the Carina-Sagittarius Arm.}
 \label{f-ANISA}
\end{center}}
\end{figure}
%%%%%%%%%%%%% Fig.2

 \subsection*{Wavelet analysis of the distribution of Cepheids}
The result of applying wavelet analysis to determine the pitch angle of two segments of spiral arms from the distribution of Cepheids on the plane ``position angle--logarithm of distance'' is shown in Fig.~\ref{f-ANISA}. To construct this figure, a total sample of 1618 Cepheids with ages in the range of 80--300 Myr was used. The size of the discrete map is $256\times256$ pixels, and the size of the map in $\theta-\ln(R/R_0)$ units is $8\times4$. The scale parameter of the wavelet transform is $c=0.1$

The parameters of the lines shown in Fig.~\ref{f-ANISA}, found using the LSM, are given in Table~\ref{t2}. In this case, the values of the wavelet map were used, satisfying the vertical values $\omega>0.3$, provided that the maps are normalized to 1 (i.e., the maximum value of the two-dimensional wavelet map along the vertical is 1.0).

%%%%%%%%%%%%%%%%%%%%%%%%%%%%%%%%%%%%%
  \begin{table*}
   \begin{center}
    \caption{
Spiral parameters found when searching for an least squares solution of a system of conditional equations of the form~(\ref{spiral-03}) using the wavelet map in Fig.~\ref{f-ANISA}
    }
   \label{t2}
   {\small
   \begin{tabular}{|l|c|c|r|r|}      \hline
 Arm             &            $a$ &             $b$ &     $i,$ deg. & $a_0$, kpc\\\hline
 II (Carina-Sagittarius)&$-0.229\pm0.008$& $-0.086\pm0.005$& $-12.88\pm0.44$&$ 7.43\pm0.05$\\
 IV  (Outer)     &$-0.222\pm0.008$& $+0.498\pm0.005$& $-12.54\pm0.50$&$13.33\pm0.06$\\
 \hline
      \end{tabular}}
     \end{center}
   \end{table*}
%%%%%%%%%%%%%%%%%%%%%%%%%%%%%%%%%%%%%

%%%%%%%%%%%%%%%%%%%%%%%% FIG.3:
\begin{figure}[t]
{ \begin{center}
   \includegraphics[width=0.82\textwidth]{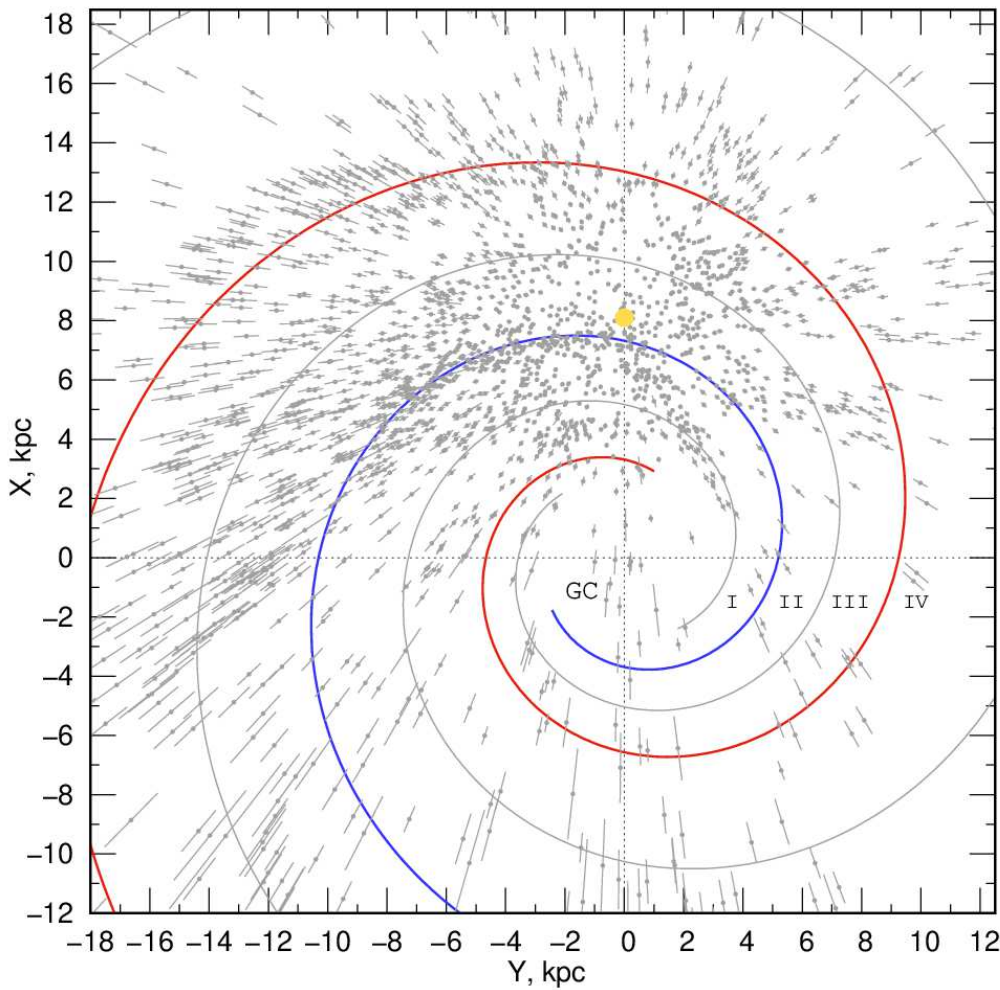}
  \caption{
The distribution of Cepheids in the projection onto the galactic $XY$ plane, the position of the Sun is marked with a yellow circle, GC~--- the center of the Galaxy, four spiral arms are shown, constructed with a pitch angle $-12.5^\circ$.}
 \label{f-XY-2214}
\end{center}}
\end{figure}

 \subsection*{Construction of a spiral pattern}
Figure~\ref{f-XY-2214} shows the distribution of all Cepheids from the catalog of Skowron et al.~(2019). The figure uses such a coordinate system in which the $X$ axis is directed towards the Sun from the center of the Galaxy, the direction of the $Y$ axis coincides with the direction of Galaxy rotation. A four-armed spiral pattern with a pitch angle of $i=-12.5^\circ$ is shown. The following segments of the spiral arms are numbered in Roman numerals: I~--- Scutum, II~--- Carina-Sagittarius (blue line), III~--- Perseus and IV~--- Outer arm (red line). Such a scale was chosen that some of the distant Cepheids remained outside the figure.

For the Carina-Sagittarius and Outer arms, the values of $a_0$ correspond exactly to those found in this paper, and for the other two arms, the values of $a_0$ are chosen ``by eye''. It can be seen from the figure that both of these arms pass well according to the data in the second and third galactic quadrants. At the same time, the Carina-Sagittarius arm is best represented, in which the densest concentration of stars in the third quadrant is clearly visible.

 \section*{CONCLUSION}\label{conclusions}
We considered a sample of classical Cepheids of the Galaxy with estimates of their distances from Skowron et al. (2019), which were calculated by these authors based on the period-luminosity relation. The main feature of this work is that the distances to the Cepheids from the work of Skowron et al. (2019) were increased by 10\% according to the results of the kinematic analysis of these Cepheids obtained by us earlier (Bobylev, and Bajkova, 2022). To determine the geometric characteristics of the Carina-Sagittarius and the Outer spiral arms, Cepheids with ages in the range of 80--300 Myr were used. The distance from the Sun to the galactic center $R_0$ in this work is assumed to be $8.1\pm0.1$~kpc.

Based on 257 Cepheids belonging to the segment of the Carina-Sagittarius arm, with ages in the range of 80--120 million years, using unit weights, $w_i=1$, when searching for the least squares solution of the system of conditional equations~(\ref{spiral-03} ), an estimate of the pitch angle of the spiral pattern $i=-12.5\pm0.5^\circ$ is obtained. The position of the center of this spiral arm on the Galaxy Center-Sun axis is found to be $a_0=7.29\pm0.05$~kpc. $i=-12.3\pm0.5^\circ$ and $a_0=13.01\pm0.06$~kpc were found from 352 older Cepheids belonging to the segment of the Outer arm, with ages of 120--300 Myr.

The calculations were repeated using weights of the form $w_i=1/\sigma^2_{\ln(R/R_0)}$.
In this case, the following estimates are obtained for the Carina-Sagittarius arm:
  $i=-12.7\pm0.4^\circ$ and $a_0=7.28\pm0.05$~kpc, and for the Outer arm
  $i=-12.0\pm0.5^\circ$ and $a_0=13.03\pm0.06$~kpc.

In this paper, we repeated the analysis by Bobylev (2022) using practically the same stars. The only difference is that the distances to the Cepheids from Skowron et al.~(2019) were increased by 10\%. We found that the lengthening of the Cepheid distance scale by 10\% changed the values of the required parameters $i$ and $a_0$ by approximately $1\sigma$.

For each segment of the spiral arm, the width parameter was found, calculated as the variance $\sigma$ with respect to the deviation from the center of the arm. For the Carina-Sagittarius arm the value of this parameter is $1\sigma=0.58$~kpc, and for the Outer arm $1\sigma=0.72$~kpc.

The calculations were repeated using a wavelet map in the ``position angle--logarithm of distance'' plane, for which a total sample of 1618 Cepheids with ages ranging from 80 to 300 Myr was used. In this case, the following estimates are obtained for the Carina-Sagittarius arm: $i=-12.9\pm0.4^\circ$ and $a_0=7.43\pm0.05$~kpc, and $i=-12.5\pm0 for the Outer arm. 5^\circ$ and $a_0=13.33\pm0.06$~kpc. We see that on the basis of many different approaches, we have obtained very similar results.

Adhering to the model of the global spiral pattern in the Galaxy with one pitch angle for all arms, we concluded that the value of this angle is close to $-12.5^\circ$.

 \bigskip\medskip{\bf REFERENCES}\medskip{\small
 \begin{enumerate}

 \item
R.I. Anderson, H. Saio, S. Ekstr\"om, C. Georgy, and G. Meynet,
Astron. Astrophys. {\bf 591}, A8 (2016).

 \item
V.V. Bobylev, Astron. Lett. {\bf 48}, 126 (2022). %i,a_0 Skowron

 \item
V.V. Bobylev, and A.T. Bajkova,
Mon. Not. R. Astron. Soc. {\bf 437}, 1549 (2014). %m=4, i=-13

 \item
V.V. Bobylev, and A.T. Bajkova, Astron. Rep. {\bf 65}, 498 (2021). %R0=8.1\pm0.1

 \item
V.V. Bobylev, and A.T. Bajkova, Astron. Rep. {\bf 66}, 545 (2022). %csale 10%

 \item
C.K. Chui, {\it Wavelets: A Mathematical Tool for Signal Analysis} (SIAM, Philadelphia, PA., 1997).

 \item
A.K. Dambis, et al., Astron. Lett. {\bf 41}, 489 (2015).

 \item
W.S. Dias, and J.R.D. L\'epine, Astrophys. J. {\bf 629}, 825 (2005).

 \item
Yu.N. Efremov, Astron. Rep. {\bf 55}, 105 (2011).

 \item
Y.M. Georgelin, and Y.P. Georgelin, Astron. Astrophys. {\bf 49}, 57 (1976).

 \item
M. Cr\'ez\'e, and M.O. Mennessier,
 Astron. Astrophys. {\bf 27}, 281 (1973).

 \item
O. Gerhard, Mem. Soc. Astron. Italiana Suppl. {\bf 18}, 185 (2011).
  %Pattern speed

 \item
E. Griv, I.-G. Jiang, and L.-G. Hou, Astrophys. J. {\bf 844}, 118 (2017).

 \item
C.J. Hao, Y. Xu, L.G. Hou, et. al.,
 Astron. Astrophys. {\bf 652}, 102 (2021).

 \item
L.G. Hou, and J.L. Han, Astron. Astrophys. {\bf 569}, 125 (2014).

 \item
L.G. Hou, and J.L. Han, Mon. Not. R. Astron. Soc. {\bf 454}, 626 (2015).

 \item
L.G. Hou, Front. Astron. Space Sci. {\bf 8}, 103 (2021).

 \item
J.R.D. L\'epine, Yu. Mishurov, and S.Yu. Dedikov,
  Astrophys. J. {\bf 546}, 234 (2001).

 \item
E.S. Levine, L. Blitz, and C. Heiles, Science {\bf 312}, 1773 (2006).

 \item
C.C. Lin, and F.H. Shu, Astrophys. J. {\bf 140}, 646 (1964).

 \item
C.C. Lin, C. Yuan, and F.H. Shu, Astrophys. J. {\bf 155}, 721 (1969).

 \item
A.M. Mel'nik, A.K. Dambis, and A.S. Rastorguev, Astron. Lett. {\bf 27}, 521 (2001).

\item
Yu.N. Mishurov, E.D. Pavlovskaia, and A.A. Suchkov, Astron. Rep. {\bf 56}, 268 (1979).

\item
Yu.N. Mishurov, I.A. Zenina, A.K. Dambis, {\it et al.},
Astron. Astrophys. {\bf 323}, 775 (1997).

\item
Yu.N. Mishurov, and I.A. Zenina, Astron. Astrophys.
{\bf 341}, 81 (1999).

 \item
A. Moitinho, R.A. V\'azquez, G. Carraro, et. al., Mon. Not. R. Astron. Soc. {\bf 368}, L77 (2006).

 \item
I.I Nikiforov, and A.V. Veselova, Astron. Lett. {\bf 44}, 81 (2018).

 \item
R. Paladini, R. Davies, and G. DeZotti, Mon. Not. R. Astron. Soc.  {\bf 347}, 237(2004).

 \item
E. Poggio, R. Drimme, T. Cantat-Gaudin,
 et al., Astron. Astrophys. {\bf 651}, 104 (2021).

 \item
M.E. Popova, and A.V. Loktin, Astron. Lett. {\bf 31}, 171 (2005).

 \item
M.J. Reid, K.M. Menten, A. Brunthaler,
   et al., Astrophys. J. {\bf 885}, 131 (2019).

 \item
D. Russeil, Astron. Astrophys. {\bf 397}, 133 (2003).

 \item
D.M. Skowron, J. Skowron, P. Mr\'oz,
  et al., Science {\bf 365}, 478 (2019).

 \item
J.H. Taylor, and J.M. Cordes, Astrophys. J. {\bf 411}, 674 (1993).

 \item
A. Udalski, M.K. Szyma\'nski, and G. Szyma\'nski,
 Acta Astron. {\bf 65}, 1 (2015). %OGLE IV

 \item
J.P. Vall\'ee, Astrophys. J. {\bf 454}, 119 (1995).

 \item
J.P. Vall\'ee, Astrophys. J. {\bf 566}, 261 (2002).

 \item
J.P. Vall\'ee, Astron. J. {\bf 135}, 1301 (2008).

 \item
J.P. Vall\'ee, New Astron. Review {\bf 79}, 49 (2017).

 \item
J.P. Vall\'ee, Astrophys. Space Sci. {\bf 363}, 243 (2018).

 \item
S. Wang, X. Chen, R. de Grijs, and L. Deng, Astrophys. J. {\bf 852}, 78 (2018).

 \item
Y. Xu, L.-G. Hou, and Y.-W. Wu,
 Research Astron. Astrophys. {\bf 18}, 146 (2018).

 \item
Y. Xu, L.G. Hou, S. Bian,
  C.J. Hao, et al.,
  Astron. Astrophys. {\bf 645}, L8 (2021).

 \end{enumerate}
 }
 \end{document}